\documentclass[notitlepage, aps,prd, reprint,preprintnumbers,onecolumn, longbibliography, showpacs]{revtex4-2}
\usepackage[%
colorlinks=true,
urlcolor=blue,
linkcolor=blue,
citecolor=blue
]{hyperref}
\usepackage{slashed, color, amsmath, amssymb, mathrsfs}
\usepackage{graphicx}
\usepackage{hyperref}
\usepackage{longtable}
\usepackage{array}
\usepackage{accents}
\usepackage{tensor}
\usepackage{longtable}
\usepackage{booktabs}
\usepackage{multirow}
\usepackage{bm}
\usepackage{subfigure}
\usepackage{epstopdf}

\graphicspath{{figures/}}
\renewcommand{\theenumi}{\roman{enumi}}

\allowdisplaybreaks

\begin{document}
%\linenumbers
\title{Origin of TeV Emission in GRB~221009A:\\
Co-effort of the External Reverse and Forward Shocks}
\author{Zhi-Lin Chen}
\author{Da-Bin Lin}\email{lindabin@gxu.edu.cn}
\author{Guo-Yu Li}
\author{En-Wei Liang}
\affiliation{Guangxi Key Laboratory for Relativistic Astrophysics, School of Physical Science and Technology, Guangxi University, Nanning, 530004, Guangxi, People’s Republic of China.}
\date{\today}
\begin{abstract}
The TeV emission detected in just five gamma-ray bursts (GRBs)
is generally ascribed to the synchrotron emission or the synchrotron self-Compton process
in the external forward shock.
The brightest gamma-ray burst, GRB 221009A, with an unprecedented detected high energy flux of TeV
emission, poses a serious challenge to the above scenario.
Different from previous works,
we involve the long bursting behavior of GRB~221009A in modeling its external-shocks.
The TeV emission together with the later multi-band afterglows of GRB~221009A
are all successfully reproduced.
It is firstly found that the TeV emission in the early phase is mainly from
the co-effort of the external reverse and forward shocks,
i.e., the inverse-Compton scattering of the synchrotron emission from the external reverse-shock by the electrons in the external forward-shock.
This is owing to that the long bursting behavior leads to a long lasting of energy injection into the external shock and the corresponding reverse-shock.
In the later phase, the TeV emission is dominated by the synchrotron self-Compton process in the external forward-shock, which is consistent with previous scenario.
Our results indicate the vital role of the external reverse-shock in shaping the early TeV emission of GRBs.
\end{abstract}
\maketitle
\section{\label{sec:Introduction}Introduction}
Gamma-ray bursts (GRBs),
featured with powerful bursts of $\gamma$-rays followed
by a long-lived afterglow emission,
are the most luminous electromagnetic explosions in the Universe.
They are typically associated with relativistic jets launched during
the collapse of a massive star or the merger of compact star binaries,
surrounded by a stellar wind or homogeneous medium, respectively \cite{2015PhR...561....1K}.
The internal dissipation process or the photosphere of the relativistic jets
are thought to release the prompt $\gamma$-rays;
and the external shocks, formed during the propagation of the jets in
the circum-burst medium,
are responsible for the long-lived afterglows \cite{1994ApJ...430L..93R,1997ApJ...490...92K,1997ApJ...476..232M,1998ApJ...497L..17S}.
Observationally and theoretically,
the prompt emission and afterglows have been intensively studied.
Very recently, B.O.A.T. (``brightest of all time'') GRB 221009A
poses a serial of debates about its prompt emission and the afterglow,
and a great many efforts have been made
\cite{2023ApJ...952L..42L,2023ApJ...949L...7F,2023ApJ...946L..31B,2023SciA....9I1405O,2023ApJ...947L..11Y,2023ApJ...947L..14Z,2024ApJ...962..115R,2024arXiv240515855B}.
Until now, a self-consistent scenario to model its TeV emission and the later multi-band afterglows
is not presented.

GRB~221009A is one of five bursts detected
with very-high-energy (VHE, $>$$\text{100\,GeV}$) $\gamma$-rays \cite{2019Natur.575..464A,2021Sci...372.1081H,2019Natur.575..459M,2024MNRAS.527.5856A}
and an unprecedentedly amount of VHE photons ($>$64000~photons) were detected by the Large High Altitude Air Shower
Observatory (LHAASO) from this burst at the early phase \cite{2023Sci...380.1390L}.
The burst is first detected by the Gamma-ray Burst Monitor (GBM) aboard Fermi satellite
at 13:16:59.99 UTC on 9 October 2022 (hereafter $T_0$) \cite{2023ApJ...952L..42L}.
The GBM records a continuous prompt emission lasting more than 600~s
and the succeeded smooth afterglow emission up to $T_0$+1464~s, in spite of $\gamma$-ray pile-up in some phases.
The prompt emission consists of two emission episodes:
a single isolated pulse from $T_0$ to $T_0$ + 20~s, i.e., the precursor;
followed by a long, extremely bright, multi-pulse emission
episode from about $T_0 + 180~{\rm s}$ to $T_0$ + 600~s, i.e., the main burst.
The long bursting behavior of GRB~2210009A
reveals a long activity of both the central engine and the launched relativistic jets.
The LHAASO detected more than 5000 VHE photons above 500~GeV within 2000~s since the GBM trigger
and the highest photon energy reaches $\sim13~{\rm TeV}$ \cite{2023SciA....9J2778C}.
The TeV light-curve presents a sharp rise at $\sim T_0$ + 230~s, turns into a relatively slow rise until the peak at $\sim T_0$ + 245~s, and then changes into a decay phase.
The smooth temporal profile of TeV light-curve contracts with the high variability of the prompt $\gamma$-rays, indicating a rather weak relation between their dissipation regions \cite{2023ApJ...952L..42L,2024arXiv240904580A}.
Thanks to the extremely high brightness of GRB 221009A,
a wealth of afterglow data from radio to GeV bands were also collected through multi-band observations \cite{2023ApJ...956L..23T,2023ApJ...946L..23L,2023NatAs...7..986B,2023ApJ...948L..12K}.
The full record of the TeV emission in the early phase and the rich of multi-band later afterglows
provide a unique opportunity to decipher the nature of VHE emission in GRBs.

A general model for the origins of VHE $\gamma$-rays is the leptonic scenario,
which is attributed to the synchrotron emission or synchrotron self-Compton process of energetic electrons accelerated by the external forward-shock \cite{2001ApJ...548..787S,2001ApJ...556.1010W,2009MNRAS.396.1163Z}.
To interpret both the TeV emission and the followed multi-band afterglows of GRB~221009A in such kind of scenario,
two special ingredients are generally used: a structured jet with a narrow core component
(half-opening angle $\sim 0.8^\circ$) and
a stratified circum-burst environment with
inner homogeneous medium surrounded by a stellar wind \cite{2023Sci...380.1390L,2024ApJ...962..115R,2024ApJ...966..141Z}.
Such kind of special jet structure and the circum-burst environment are very unusual,
e.g.,
the characteristic half-opening angle of the jet inferred based on the observed jet break time
in afterglow is $\sim3.6^\circ$ \cite{2001ApJ...562L..55F}.
In fact, the narrow jet core and the inner homogeneous medium were proposed to
interpret the light-curve morphology of early TeV emission.
We note that the main burst of GRB~221009A
overlaps on the early TeV emission,
implying a continuous energy injection into the external shock in situ or subsequently.
Correspondingly, a long-lasting external reverse-shock appears and accelerates the electrons therein.
The continuous energy injection and the arose external reverse-shock should modify the emission of the external shock.
In this letter, we model the TeV emission together with the later multi-band afterglows
by involving the fact that a long-lasting energy injection into the external shock appears in GRB~221009A owing to its long bursting behavior.

\section{Methods}\label{sec:Methods}
\subsection{Reproduction of the jet power history}
The central engine of GRB~221009A continuously drives relativistic collimated jets.
The propagation of the early launched jet in the stellar wind results to the formation of the external shocks.
The later launched jets catch up with the formed external shock and thus
inject energy into the external shock through the reverse-shock.
The history of the central engine and its launched jets are crucial.
In our model, the power $P_{\rm jet}$ of the jet driven from the central engine is prescribed
based on Fermi-GBM observations,
i.e.,
$P_{\rm jet}=L_{\gamma, \rm iso}/\eta_{\gamma}$,
where $L_{\gamma, \rm iso}$ is the isotropic luminosity of the prompt emission in 1\,keV-10\,MeV energy range
and $\eta_{\gamma}$ is the radiation efficiency of the corresponding jet.
Thus, the residual energy in the jet is $P_{\rm jet}(1-\eta_{\gamma})$.

$L_{\gamma, \rm iso}$ is estimated as follows.
We obtain daily data covering the time range of GRB~221009A from the Fermi-GBM public data
archive.
Based on the position history file,
we select two sodium iodide detectors (n4 $\&$ n8) and one bismuth germanium oxide detector (b1) with optimal viewing angle for spectral analysis.
Since the continuous spectroscopy response matrices provided by the Fermi GBM Trigger Catalog only covers $\sim 125\,{\rm s}$ before Fermi triggered and $\sim 480\,{\rm s}$ after Fermi triggered,
we use the GBM Response
Generato ($\rm SA\_GBM\_RSP\_Gen.pl$) to produce the response files for n4, n8, and b1 detectors during [$T_{\rm 0} - 100\,{\rm s}$, $T_{\rm 0} + 2000\,{\rm s}$],
where the source location of (RA, Dec)= ($288.26^{\circ}$, $19.77^{\circ}$) is used.
The GBM Data Tools
is used to extract the light-curve based on the Time Tagged Event (TTE) data.
We use a polynomial algorithm to fit background with order of 3,
and the observations in the periods of $(T_{\rm 0}-140.0{\rm s},T_{\rm 0}-100.0{\rm s}), (T_{\rm 0}-80.0{\rm s},T_{\rm 0}-40.0{\rm s}), (T_{\rm 0}+60.0{\rm s},T_{\rm 0}+110.0{\rm s}), (T_{\rm 0}+1600.0{\rm s}, T_{\rm 0}+1640.0{\rm s}), (T_{\rm 0}+1700.0{\rm s}, T_{\rm 0}+1740.0{\rm s})$,
and $(T_{\rm 0}+1760.0{\rm s}, T_{\rm 0}+1790.0{\rm s})$ are selected as the background.
The time-resolved radiation spectrum per $1.024\,{\rm s}$ is fitted with a Band function to obtain the 1keV-10MeV energy flux $F_{\gamma}$
and thus $L_{\gamma, \rm iso}=4\pi D_{\rm L}^2 g_{\rm c}F_{\gamma}$ is obtained,
where $D_{\rm L}=10^{27.39}\,{\rm cm}$ is the luminosity distance of GRB~221009A and $g_{\rm c}$ is the correcting factor of the energy flux described as follows.
%%%%%%%%%%%%%%%%%%%%%%%%%%%%%%%%%
\begin{figure}
\includegraphics[width=0.5\textwidth]{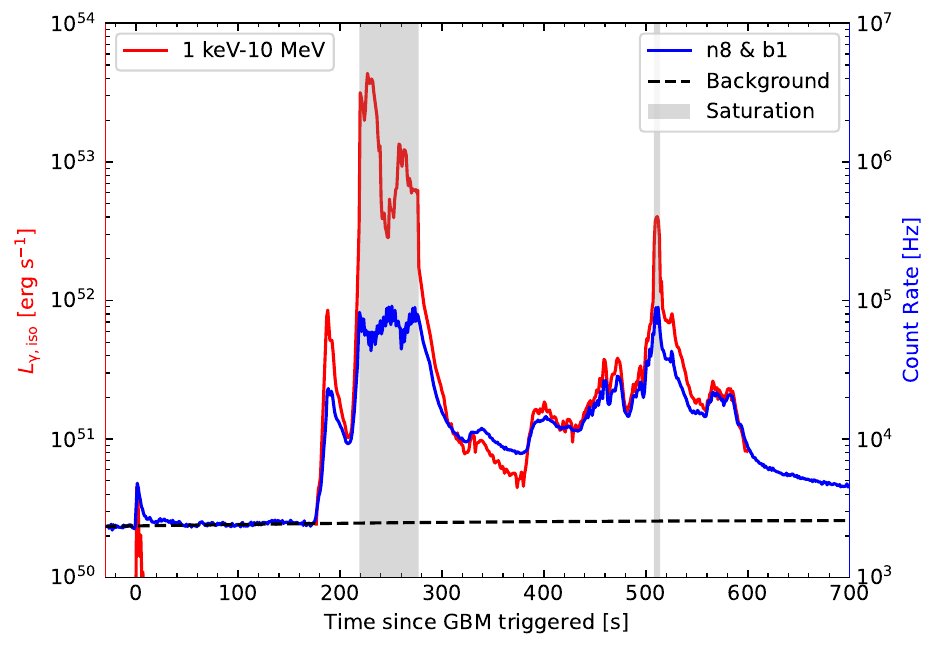}
\caption{\textbf{The estimated isotropic luminosity $L_{\gamma, \rm iso}$ of the prompt emission (red line) based on Fermi-GBM observations.}
Here, the count rate from n8 $\&$ b1 detectors of Fermi-GBM is showed with blue line
and the correction on the energy flux during the BTI of Fermi-GBM is based on the comparison between the GECAM data and Fermi-GBM data.}
\label{fig2-PeomptEmission}
\end{figure}
%%%%%%%%%%%%%%%%%%%%%%%%%%%%%%%%%

During the Bad Time Intervals (BTI) of Fermi-GBM for GRB~221009A,
i.e., [219, 277]s and [508, 514]s,
the obtained energy flux $F_{\gamma}$ may not really reflect the energy flux.
Then, we introduce a correcting factor $g_{\rm c}$.
Besides the BTI of Fermi-GBM, $g_{\rm c}=1$ is set.
Within the BTI of Fermi-GBM, the value of $g_{\rm c}$ is set as follows.
During the BTI of Fermi-GBM, the Gravitational-wave high-energy Electromagnetic Counterpart All-sky Monitor (GECAM)
is free of $\gamma$-ray pile-up.
With GECAM, Ref. \cite{2023arXiv231011821W} has reported
the energy flux $\hat{F}_{\gamma}$ in four time intervals of 231-240~s, 240-248~s, 248-326~s, and 326-650~s, i.e., $\hat{F}_{\gamma, 231-240}=3.66 \times 10^{-3} \rm \,erg/cm^{2}/s$,
$\hat{F}_{\gamma, 240-248}=7.18 \times 10^{-4}\,\rm erg/cm^{2}/s$,
$\hat{F}_{\gamma, 248-326}=4.37 \times 10^{-4}\,\rm erg/cm^{2}/s$,
and $\hat{F}_{\gamma, 326-650}=4.67 \times 10^{-5}\,\rm erg/cm^{2}/s$, respectively.
Then, we also estimate the energy flux $F_{\gamma}$ in these four time intervals based on spectral analysis method and the observations of Fermi-GBM.
The results are $F_{\gamma, 231-240}=4.32 \times 10^{-4}\,\rm erg/cm^{2}/s$, $F_{\gamma, 240-248}=2.04 \times 10^{-4}\,\rm erg/cm^{2}/s$, $F_{\gamma, 248-326}=1.32 \times 10^{-4}\,\rm erg/cm^{2}/s$, and $F_{\gamma, 326-650}=2.34 \times 10^{-5}\,\rm erg/cm^{2}/s$, respectively.
Then, $g_{\rm c}=\hat{F}_{\gamma, 231-240}/F_{\gamma, 231-240} \sim 8.47$,
$\hat{F}_{\gamma, 240-248}/F_{\gamma, 240-248} \sim 3.51$,
$\hat{F}_{\gamma, 248-326}/F_{\gamma, 248-326} \sim 3.31$,
and
$\hat{F}_{\gamma, 326-650}/F_{\gamma, 326-650} \sim 1.99$
are set for the time intervals of
219-240\,s,
240-248\,s,
248-277\,s,
and 508s-514\,s, respectively.
The obtained $L_{\gamma, \rm iso}$ based on Fermi-GBM observations is shown in Fig.\ref{fig2-PeomptEmission}, and the spectral analysis resuts are reported in Table~\ref{tab2}.
To simplify, the radiation efficiency $\eta_{\gamma}$ of the jet is assumed to be the same
during each time interval of 214-224\,s, 224-252\,s, 253-326\,s, 326-450\,s, and 326-600\,s, respectively.
The bulk Lorentz factor of the jets in these time intervals is set as the same $\Gamma_{\rm jet}$
with the same half-opening angle $\theta_{\rm jet}$.

%%%%%%%%%%%%%%%%%%%%%%%%%%%%%%%%
\begin{table}[h]
\caption{Spectral analysis results of the prompt emission.}
\label{tab2}%
\begin{tabular}{@{}lcccc@{}}
\toprule
Time bin[s]& $\alpha$  & $\beta$ & $\lg\left ( \frac{E_{\rm cut}}{\mathrm{keV}} \right ) $ & $\lg\left ( \frac{\rm Flux}{\rm  erg/cm^{2}/s } \right ) $\\
\midrule
180-210 & $-1.08^{+0.01}_{-0.01}$ & $-2.46^{+0.02}_{-0.02}$ & $2.87^{+0.01}_{-0.01}$ & $-4.51^{+0.01}_{-0.01}$ \\
210-218  & $-1.13^{+0.01}_{-0.01}$ & $-2.18^{+0.02}_{-0.02}$ & $3.04^{+0.01}_{-0.01}$ & $-4.15^{+0.01}_{-0.01}$ \\
218-231 \footnotemark[1] & $-0.92^{+0.01}_{-0.01}$ & $-2.40^{+0.01}_{-0.01}$ & $2.84^{+0.02}_{-0.02}$ & $-2.34$ \\
231-240 \footnotemark[2] & $-0.92^{+0.01}_{-0.01}$ & $-2.40^{+0.01}_{-0.01}$ & $2.84^{+0.02}_{-0.02}$ & $-2.44^{+0.01}_{-0.01}$ \\
240-248 \footnotemark[2] & $-1.19^{+0.01}_{-0.01}$ & $-2.72^{+0.03}_{-0.03}$ & $2.75^{+0.02}_{-0.02}$ & $-3.14^{+0.02}_{-0.01}$ \\
248-326 \footnotemark[2] & $-1.15^{+0.01}_{-0.01}$ & $-2.38^{+0.01}_{-0.01}$ & $2.98^{+0.01}_{-0.01}$ & $-3.36^{+0.01}_{-0.01}$ \\
326-375  & $-1.21^{+0.03}_{-0.03}$ & $-2.15^{+0.01}_{-0.01}$ & $1.69^{+0.02}_{-0.02}$ & $-5.06^{+0.01}_{-0.01}$ \\
375-430  & $-1.60^{+0.01}_{-0.01}$ & $-2.45^{+0.03}_{-0.03}$ & $2.84^{+0.01}_{-0.01}$ & $-4.87^{+0.01}_{-0.01}$ \\
430-480  & $-1.52^{+0.01}_{-0.01}$ & $-2.50^{+0.02}_{-0.02}$ & $2.80^{+0.01}_{-0.01}$ & $-4.62^{+0.01}_{-0.01}$ \\
480-550 & $-1.39^{+0.01}_{-0.01}$ & $-2.42^{+0.01}_{-0.01}$ & $2.96^{+0.01}_{-0.01}$ & $-4.22^{+0.01}_{-0.01}$ \\
550-600 & $-1.59^{+0.01}_{-0.01}$ & $-2.41^{+0.01}_{-0.01}$ & $2.66^{+0.01}_{-0.01}$ & $-4.73^{+0.01}_{-0.01}$ \\
\botrule
\end{tabular}
\footnotetext[1]{There is no spectral analysis in this period from Ref. \cite{2023arXiv231011821W}. The radiation spectrum of this period is taken as the same as that of 231-240\,s but involving the correction on the flux.}
\footnotetext[2]{The radiation spectra during the BTI of Fermi-GBM is taken from Ref. \cite{2023arXiv231011821W}.}
\end{table}
%%%%%%%%%%%%%%%%%%%%%%%%%%%%%

\subsection{Model of the external shock}\label{Sec_ES}
The circum-burst environment of GRB~221009A is the blew-off stellar wind from the massive star \cite{2023SciA....9I1405O,2023ApJ...946L..23L,2024NatAs...8..774B}.
The particle density of wind at different radii $r$ is given by $n(r)=3\times10^{35}r^{-2}A_{\star}\rm{~cm}^{-1}$
with the wind parameter $A_{\star}=\dot{M}/10^{-5}~M_{\odot}\ \rm{yr}^{-1}\left(4\pi \upsilon_{w}/1000~\rm{km}~\rm{s}^{-1}\right)^{-1}$,
where $\dot{M}$ is the mass-loss rate of the massive star and $\upsilon_{w}$ is the wind velocity.
According to Fig.~\ref{fig2-PeomptEmission},
the central engine of GRB~221009A is active during 0-20\,s (precursor),
dies out in 20-180~s, and revives in 180-650\,s (the main burst).
The precursor of the central engine is very weak, compared with the first pulse ($\sim$180-210\,s) of the main burst.
In addition, the afterglows after $\sim$230\,s,
e.g., the TeV emission and the later multi-band afterglows, are the main focus of this paper.
Then, we take the jet produced the first pulse as the initial fireball that drives the external shocks.
The isotropic kinetic energy of the initial fireball is taken as $E_{\rm{k, 0}}=1.8\times 10^{53}$\,erg, corresponding to a typical $\gamma$-rays production efficiency of $30\,\%$, and we also adopt the typical value for its bulk Lorentz factor, i.e., $\Gamma_{0}=200$.
The adopted parameters of the
initial fireball do not affect our modelling on the TeV emission and the later multi-band afterglows
since there is strong energy injection into the external shock subsequently.
In addition, the very early afterglows based on our initial fireball are also consistent with the observed flux limits.

The initial fireball is decelerated during its propagation in the stellar wind.
Observationally, GRB~221009A enters a strong bursting after the first pulse of the main burst.
There is a rapid rise of the isotropic luminosity from $\sim210\,{\rm s}$ and
the Fermi-GBM detectors are $\gamma$-ray pile-up after a short time ($\sim$9\,s).
The strong bursting behavior in this phase leads to an enormous energy injection into the external shock in situ or subsequently, which is consistent with the rapid rise of the TeV onset at around 230\,s.
In our modeling, the later launched jets, responsible for the pulses in the period of 210-600\,s,
would catch up and collide with the decelerated initial fireball at certain times.
We take the minimum time for the later launched jets to collide with the decelerated initial fireball as $T_{\rm min}$.
Since the values of $E_{\rm{k, 0}}$ and $\Gamma_{0}$ can not be well inferred,
the value of $T_{\rm min}$ is taken as a free parameter in our fitting.

During the development of the forward/reverse shock,
the electrons are accelerated and the magnetic fields are amplified in the shock.
The fractions of the shock internal energy used to accelerate non-thermal electrons and amplify the magnetic fields
are generally set as $\varepsilon_{\rm e, fs}$ ($\varepsilon_{\rm e,rs}$) and $\varepsilon_{B, \rm fs}$ ($\varepsilon_{B, \rm rs}$), respectively.
Here, the subscript ``$_{...,\rm fs}$'' and ``$_{...,\rm rs}$'' represent the parameters in the forward and reverse shocks, respectively.
The energy spectrum $dN_{\rm e}/d\gamma_{\rm e}$ of the accelerated electrons is generally assumed as a power-law distribution,
i.e., $dN_{\rm e,  fs}/d\gamma_\mathrm{e}=Q_{\rm fs}(\gamma_{\rm e}/\gamma_{\rm e, m,\rm fs})^{-p_{\rm fs}}$
or $dN_{\rm e,  rs}/d\gamma_\mathrm{e}=Q_{\rm rs}(\gamma_{\rm e}/\gamma_{\rm e, m,\rm rs})^{-p_{\rm rs}}$,
where $\gamma_{e}$ is the electron's Lorentz factor,
$\gamma_{\rm e, m, fs}$ ($\gamma_{\rm e, m, rs}$) is the minimum Lorentz factor of the accelerated electrons,
and $p_{\rm fs}$ ($p_{\rm rs}$) is the power-law index.
The values of $Q_{\rm fs}$ ($Q_{\rm rs}$) and $\gamma_{\rm e,m,fs}$ ($\gamma_{\rm e,m,rs}$) are  obtained by solving the equations about the total energy and total number of the non-thermal electrons \cite{2008MNRAS.384.1483F,2018ApJS..234....3G}.
Since the electrons swept by the shocks are not all accelerated \cite{2005ApJ...627..861E,2008ApJ...682L...5S},
we take the fraction of electrons accelerated to the non-thermal distribution in the total electrons as $\xi_{e,\rm fs}$ or $\xi_{e,\rm rs}$.
In the forward shock, the minimum and maximum Lorentz factor of the accelerated electrons \cite{1998ApJ...497L..17S,2012MNRAS.427L..40K} are respectively
$\gamma _{\rm e,m,fs}=\Gamma _{\rm fs}\,\varepsilon _{\rm e,fs}( p_{\rm fs}-2  ) m_{\rm p}/[( p_{\rm fs}-1 ) m_{\rm e}\zeta_{e,\rm fs}]$
and
$\gamma _{\rm e,max,fs}=\sqrt{9m^2_{\rm e}c^4/\left ( 8 B_{\rm fs} q^3_{\rm e} \right ) } $,
where the magnetic field is $B_{\rm fs}= \left ( 32 \pi \rho_{\rm fs}\,\varepsilon _{B,{\rm fs}} \right )^{1/2} \Gamma_{\rm fs} c$ with the mass density $\rho_{\rm fs} =m_{\rm p}n ( r_{\rm fs}) $,
$r_{\rm fs}$ is the distance of the forward shock with respect to the central engine,
$\Gamma_{\rm fs}$ is the bulk Lorentz factor of the post-shock medium,
and $m_{\rm e}$, $m_{\rm p}$, $q_{\rm e}$, and $c$ are the electron mass, proton mass, electron charge, and light speed, respectively.
For the reverse shock, the shock Lorentz factor is $\Gamma_{\rm rs}=\Gamma_{\rm jet}\Gamma_{\rm fs}( 1-\beta_{\rm jet} \beta_{\rm fs}) $ with $\beta_{\rm fs}=\sqrt{1-1/\Gamma_{\rm fs}^2}$ and  $\beta_{\rm jet}=\sqrt{1-1/\Gamma_{\rm jet}^2}$.
Analogy to the forward shock, one can infer the post-shock magnetic field and non-thermal electrons for the reverse shock based on the same equation but replacing the subscript ``$_{...\rm fs}$'' with ``$_{...\rm rs}$'',
where $\rho_{\rm rs} =P_{\rm jet} ( 1-\eta _{\gamma } )/(4\pi r_{\rm rs}^2\Gamma_{\rm jet}c^{2})$.
When the energy injection ceases, the evolution of the magnetic fields in the post-reverse shock region is assumed to be regulated by the energy conservation.
The complete dynamics of the external shocks involving the energy injection through the reverse shock is estimated based on \cite{2013MNRAS.433.2107N}.

The non-thermal electrons suffer from the cooling due to the synchrotron emission and the inverse-Compton process \cite{1965PhRv..137.1306J,1970RvMP...42..237B}.
The inverse-Compton process of the electrons upscatters the seed photons to the VHE photons,
where the synchrotron emission of the electrons in both the forward and reverse shocks
and the soft photons of the prompt emission serve as the seed photons.
The radiation spectra of the prompt emission listed in Table~\ref{tab2} are used to calculate
the soft photons during the prompt emission phase.
In our calculations, the synchrotron self-absorption, the Klein-Nishina effect on the inverse-Compton process,
and the absorption of VHE photons due to the two-photon pair production are all considered \cite{rybicki1991radiative,1967PhRv..155.1408G,2011ApJ...732...77M}.

\section{\label{sec:Result}Result}

\begin{figure}
\includegraphics[width=\columnwidth]{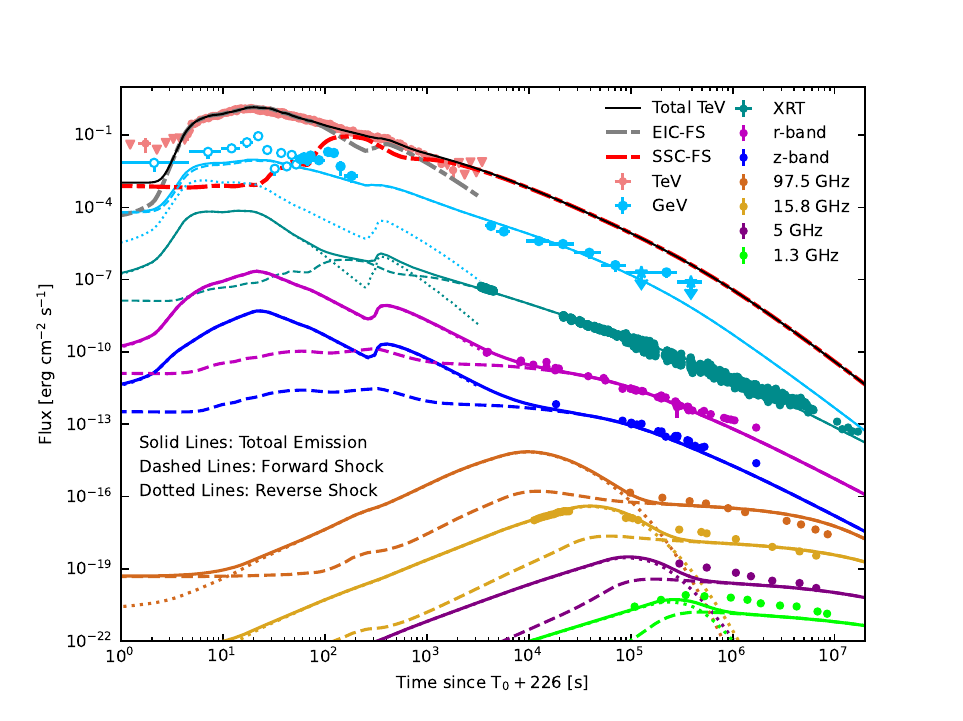}
\caption{\textbf{Modeling on the multi-waveband afterglows of GRB~221009A.}
The total emission from both the forward and reverse shocks of our model is represented with the solid lines,
and the contribution from the forward (reverse) shock is depicted with dashed (dotted) lines.
The two main contributors for the early TeV emission,
i.e., the inverse-Compton of the synchrotron emission from the reverse shock by the electrons in the forward shock (EIC-FS) and the synchrotron self-Compton process in the forward shock (SSC-FS),
are shown with gray and red thick dashed-dotted lines, respectively.
All the data and the corresponding modeling light curves are depicted with the same color,
and shifted by multiplying the factor of $10^5$, $10^3$, $10^0$, $10^{-1}$, $10^{-2.5}$, $10^{-2}$, $10^{-2.5}$, $10^{-3.5}$, and $10^{-4}$ from top to bottom, respectively.
} \label{fig2-lightcurves}
\end{figure}
%%%%%%%%%%%%%%%%%%%%%%%%%%%%%%%%%%%%%%%%%%
\begin{figure}
\includegraphics[width=\columnwidth]{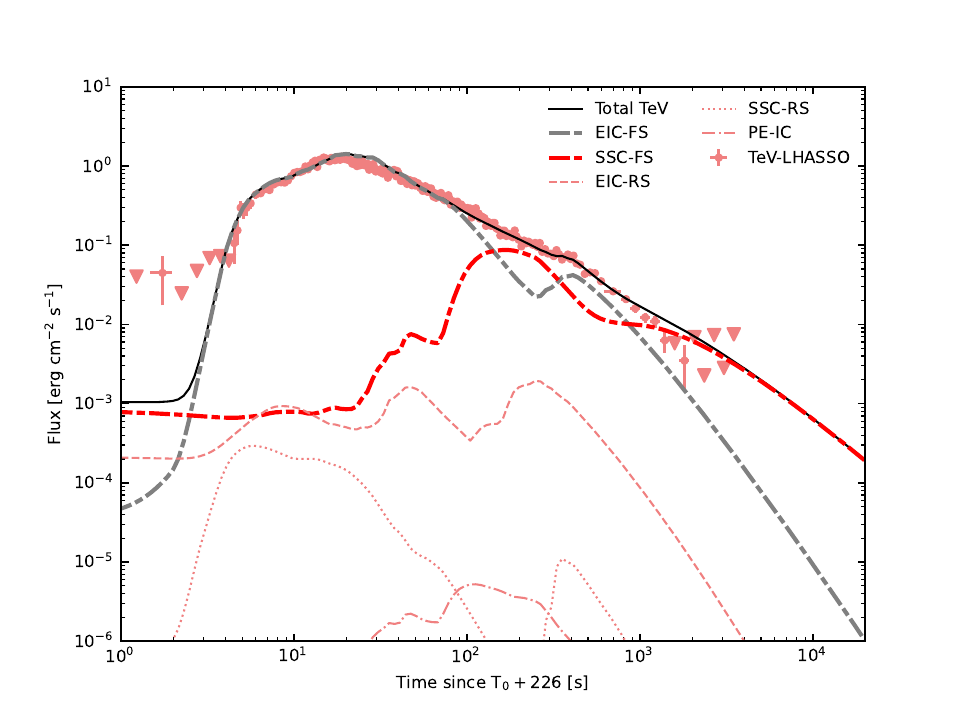}
%\captionsetup{justification=raggedright, singlelinecheck=false}
 \caption{{\bf Zoom-in on the TeV light-curve modeling with all contributors}.
Except the contributors in Fig.~\ref{fig2-LightCurve-TeVGeV},
the inverse-Compton of the synchrotron emission from the forward shock by the electrons in the reverse shock (EIC-RS), the synchrotron self-Compton process in the reverse shock (SSC-RS), and the inverse-Compton of the prompt soft photons by the electrons in the external shocks (PE-IC) are also plotted.} \label{fig2-LightCurve-TeVGeV}
 \end{figure}
%%%%%%%%%%%%%%%%%%%%%%%%%%%%%%%

Fig.~\ref{fig2-lightcurves} shows our modeling on the TeV emission together with the contemporaneous GeV emission and the later X-ray-optical-radio afterglows,
where the solid lines are the results from our model
and the parameters of our model are reported in Table~\ref{tab1}.
Here, the intrinsic TeV flux data corrected for the EBL attenuation
is taken from LHAASO \cite{2023Sci...380.1390L},
the GeV data is from Fermi-LAT \cite{2024arXiv240904580A} (and the corrected GeV data during BIT is showed  with an empty circle),
the X-ray data comes from the Swift-XRT,
the optical data \cite{2023ApJ...948L..12K} is corrected for the Galactic and host galaxy extinction \cite{2011ApJ...737..103S,1992ApJ...395..130P},
and the radio data is collected from \cite{2023ApJ...946L..23L,2023NatAs...7..986B}.
One can find that the total emission (solid lines) from both the external forward and reverse shocks
well fits the multi-band afterglows.
Importantly, the light-curve morphology of TeV emission,
including the first sharp rise, the followed slow rise to the peak,
and the slow decay, is well-reproduced based on a normal jet (opening angle $\theta_{\rm jet}=6.93^\circ$) propagating in a stellar wind ($A_*=2.8$).
Please see Fig.~\ref{fig2-LightCurve-TeVGeV} for the zoom-in on the TeV light-curve modeling.
The light-curve break of TeV emission at around 730\,s is formed owing to the end of the energy injection into the external shock in this phase.
Importantly, there are two main contributors for the TeV emission,
i.e.,
the inverse-Compton of the synchrotron emission from the reverse shock (EIC-FS)
and that from the forward shock (SSC-FS) by the electrons in the forward shock.
These two contributors dominate the TeV emission by turn,
with EIC-FS dominating the very early TeV emission ($\sim$230-1000\,s)
and SSC-FS dominating the later TeV emission ($\gtrsim$1000\,s).
This is the first burst that the external reverse shock acts vital role
in shaping the TeV emission.

%%%%%%%%%%%%%%%%%%%%%%%%%%%%%%%%%%%%%%%%%%
\begin{figure}
\includegraphics[width=\columnwidth]{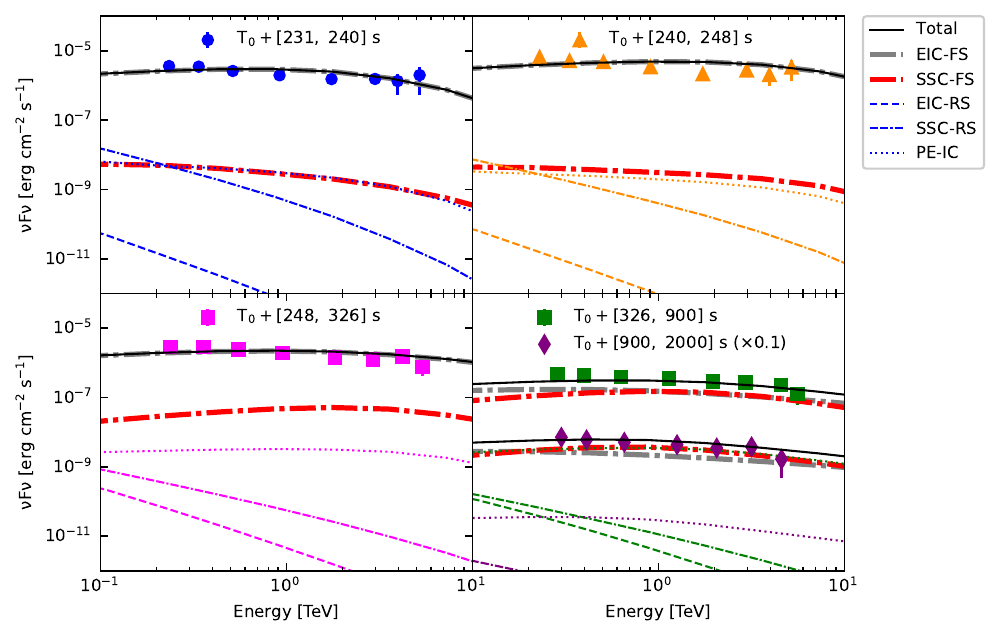}
\caption{TeV radiation spectra during the periods of 231-240\,s, 240-248\,s,
248-326\,s, 326-900\,s, and 900-2000\,s,
where the data points are the intrinsic TeV emission from LHAASO \cite{2023Sci...380.1390L}
and the solid lines are the total emission from our model.
The EIC-RS, SSC-RS, and PE-IC are plotted with
the same color as that of the data.} \label{fig2-spectrum}
\end{figure}

The spectra in the TeV energy range for different phases are also presented in Fig.~\ref{fig2-spectrum}.
As indicated in Fig.~\ref{fig2-spectrum},
the contribution from the PE-IC process is negligible for the TeV emission.
During the rising to the peak of TeV emission, e.g., 231-240\,s and 240-248\,s,
it is found that the TeV spectra is softer than the SSC spectrum \cite{2023Sci...380.1390L}.
This can also be found by comparing the TeV spectrum in this phase with the red dashed-dot-dotted lines in Fig.~\ref{fig2-spectrum}.
In our model,
the TeV emission in this phase is dominated by the EIC-FS.
This is owing to the strong energy injection and thus an abundant of soft photons released in the reverse shock in this phase.
In the decay phase of TeV emission, however,
the energy injection becomes weak and thus the soft photons from the reverse shock is lessen.
Corresponding, the SSC-FS becomes dominant, which is consistent with the TeV spectra in this phase, e.g., 326-2000\,s.
The spectral evolution that appears across the entire TeV afterglow can be well reproduced
under the co-effort of the external reverse and forward shocks.

\section{\label{sec:Discussion_and_conclusions}Conclusion}
Compared with the external forward shock,
the external reverse shock is generally found to dominate the emission in the low-energy regime (e.g., optical band) during the early phase of GRBs' afterglow.
Then, the reverse shock would play a vital role in shaping the early TeV emission produced by
the inverse-Compton process.
GRB~221009A is the first burst that witnesses the co-effort of the external reverse and forward shocks on the formation of the very-high-energy $\gamma$-rays in GRBs.
It is showed that the inverse-Compton of the synchrotron emission from the reverse shock by the electrons in the forward shock is responsible for the early TeV afterglow.
The synchrotron self-Compton process in the forward shock is responsible for the later TeV afterglow, consistent with the previous studies.
In the future, the TeV data as an indicator for the afterglow onset and the other multi-band data during the same phase may be utilized to study the particle acceleration in the forward/reverse shock, the jet properties, and the circum-burst environment.

%%%%%%%%%%%%%%%%%%%%%%%%%%%%%%%

\begin{table*}[h]
\caption{Value of parameters in our model, associated with the external forward/reverse shock, the later launched jets, and the circum-burst environment.}\label{tab1}%
\begin{tabular}{@{}llll@{}}
\toprule
{} & Physical parameters  & Symbols & Values\\
\midrule
Forward shock & Electron spectral index   & $p_{fs}$  & $2.55$  \\
{}    & Electron energy fraction   & $\varepsilon_{e,\rm fs}$  & $10^{-2.57}$  \\
{}    & Non-thermal electron fraction   & $\xi_{e,\rm fs}$  & $10^{-2.25}$  \\
{}    & Magnetic energy fraction   & $\varepsilon_{b,\rm fs}$  & $10^{-4.91}$  \\
%\midrule
Reverse shock & Electron spectral index   & $p_{\rm rs}$  & $3.51$  \\
{}    & Electron energy fraction   & $\varepsilon_{e,\rm rs}$  & $10^{-2.19}$  \\
{}    & Non-thermal electron fraction   & $\xi_{e,\rm rs}$  & $10^{-2.10}$  \\
{}    & Magnetic energy fraction   & $\varepsilon_{b,\rm rs}$  & $10^{-3.64}$  \\
%\midrule
Later launched jets   & Radiation efficiency\footnotemark[1]   & $\eta_{\gamma}$  & $\left [ 10^{-1.21}, 10^{-1.23}, 10^{-0.62}, 10^{-0.55}, 10^{-1.59} \right ]$ \\
{}    & Lorentz factor   & $\Gamma_{\rm jet}$  & $10^{2.76}$  \\
%\midrule
{}    & Jet half-opening angle   & $\theta_{\rm jet}$  & $10^{-1.21}\,{\rm rad}$  \\
{}    & Minimum injection time   & $T_{\rm min}$  & $222.7\,{\rm s}$  \\
Environment    &  Stellar wind & $A_*$  & $2.88$  \\
\midrule
Initial fireball\footnotemark[2]    & Isotropic kinetic energy   & $E_{\rm k,iso}$  & ${10^{53.2}\,{\rm erg}}$  \\
{}    & Initial bulk Lorentz factor   & $\Gamma_{0}$  & 200  \\
\botrule
\end{tabular}
\footnotetext[1]{The later launched jets are responsible for the prompt emission during 210-600\,s.
This period is divided into 5 time bins, i.e., 210-224\,s, 224-252\,s, 253-326\,s, 326-450\,s, and 450-600\,s. The radiation efficiency is estimated in each time bin and the luminosity of the prompt emission is estimated in $1\,{\rm keV}$-$10\,{\rm MeV}$  energy range.}
\footnotetext[2]{For the adopted value of the initial fireball parameters, see Section~\ref{Sec_ES} for details.}
\end{table*}

\acknowledgements
We acknowledge the use of the Fermi and Swift archive's public data.
We thank Jia Ren  and Shan-Shan Weng for helpful discussion and Xing Yang for helpful in computing performance.
This work is supported by the National Natural Science Foundation of China (grant Nos. 12273005 and 12133003) and the National Key R\&D Program of China (grant No. 2023YFE0117200).

\clearpage

\bibliography{ms}

\end{document}